\documentclass[MSNbibl,nameyear,dvips]{arxstspdf}
\usepackage{graphicx}
\usepackage{flushend}
\usepackage{stfloats}

\volume{28}
\issue{3}
\pubyear{2013}
\firstpage{283}
\lastpage{288}
\doi{10.1214/13-STS438} 

\begin{document}
\begin{frontmatter}

\title{The True Title of Bayes's Essay}
\runtitle{The True Title of Bayes's Essay}

\begin{aug}
\author[a]{\fnms{Stephen M.} \snm{Stigler}\corref{}\ead[label=e1]{stigler@uchicago.edu}}
\runauthor{S.~M. Stigler}

\affiliation{University of Chicago}

\address[a]{Stephen M. Stigler is the Ernest DeWitt Burton Distinguished Service Professor in the Department of
Statistics, University of Chicago, 5734 University Avenue, Chicago, Illinois 60637, USA \printead{e1}.}

\end{aug}

\begin{abstract}
New evidence is presented that Richard Price gave Thomas Bayes's famous essay a
very different title from the commonly reported one. It is argued that this implies Price almost
surely and Bayes not improbably embarked upon this work seeking a defensive tool to combat David
Hume on an issue in theology.
\end{abstract}

\begin{keyword}
\kwd{Thomas Bayes}
\kwd{Richard Price}
\kwd{Bayes's theorem}
\kwd{history}
\end{keyword}

\end{frontmatter}
\begin{figure}

\includegraphics{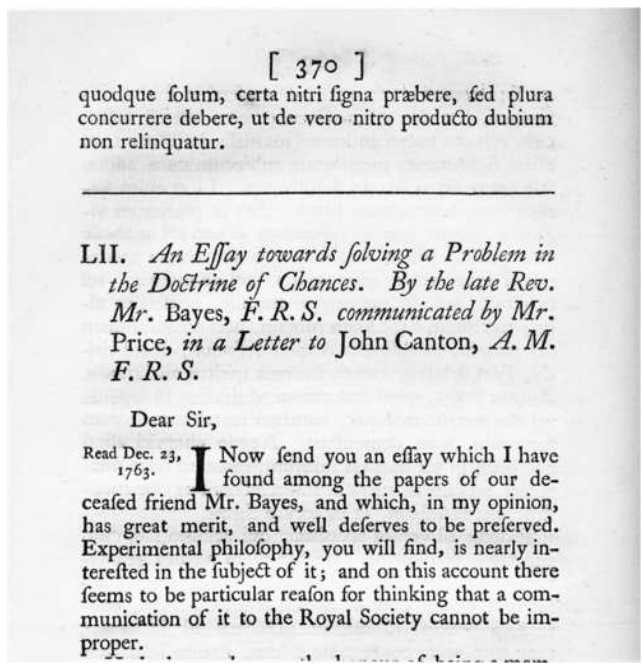}

\caption{The heading for Bayes
(\citeyear{Bay64}).}\label{fig1}\vspace*{6pt}
\end{figure}
\begin{figure}[t]

\includegraphics{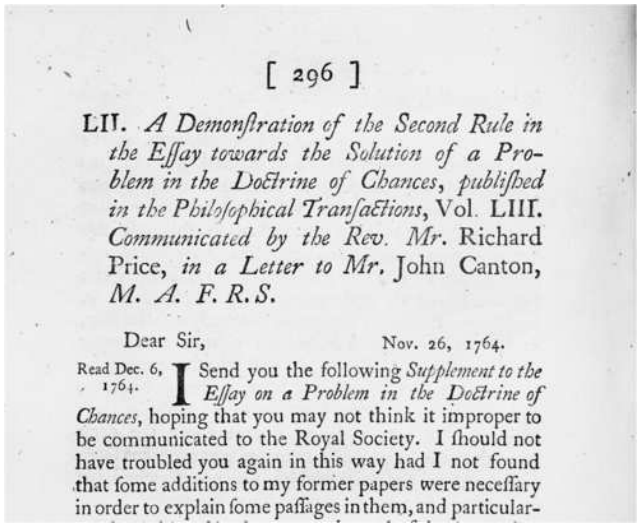}

\caption{The heading for Price (\citeyear{Pri65}).}\label{fig2}
\end{figure}

Monday 23 December 2013 is the 250{th} anniversary of the date Richard Price presented Thomas
Bayes's famous paper at a meeting of the Royal Society of London. The paper was published in 1764 as
part of the 1763 volume of the \textit{Philosophical Transactions} of the Royal Society, with the
block of print shown in Figure~\ref{fig1} at its head. In December 1764 Richard Price read a follow-up paper
with himself as author (Figure~\ref{fig2}); it was published in 1765 as part of the volume for 1764. All
modern readers have taken these article heads as the titles of the papers; the first as ``An Essay
toward solving a Problem in the Doctrine of Chances;'' the second as ``A Demonstration of the
Second Rule in the Essay toward the Solution of a Problem in the Doctrine of Chances.'' But Richard
Price (and perhaps Bayes as well) had very different titles in mind.

At that time, it was the occasional practice of the Royal Society to supply authors with offprints
of published papers, generally before the appearance of the printed volume, based upon the same
print block used for the \textit{Transactions} but with the pagination beginning with the number
1 and the first page from the journal version set to accommodate the different format. Presumably
this was only done when the author requested and at the author's expense. The offprints were
supplied with a cover page. In Bayes's case the offprints produced in 1764 had a cover page showing
a dramatically different title:

\begin{quote}
A Method of Calculating the Exact Probability of All Conclusions founded on Induction.
\end{quote}

The journal title was retained on page 3 of the offprint, as a subtitle. A year later, in 1765,
offprints of the second paper were produced with the title:

\begin{quote}
A Supplement to the Essay on a Method of Calculating the Exact Probability of All Conclusions
founded on Induction.
\end{quote}

\begin{figure}

\includegraphics{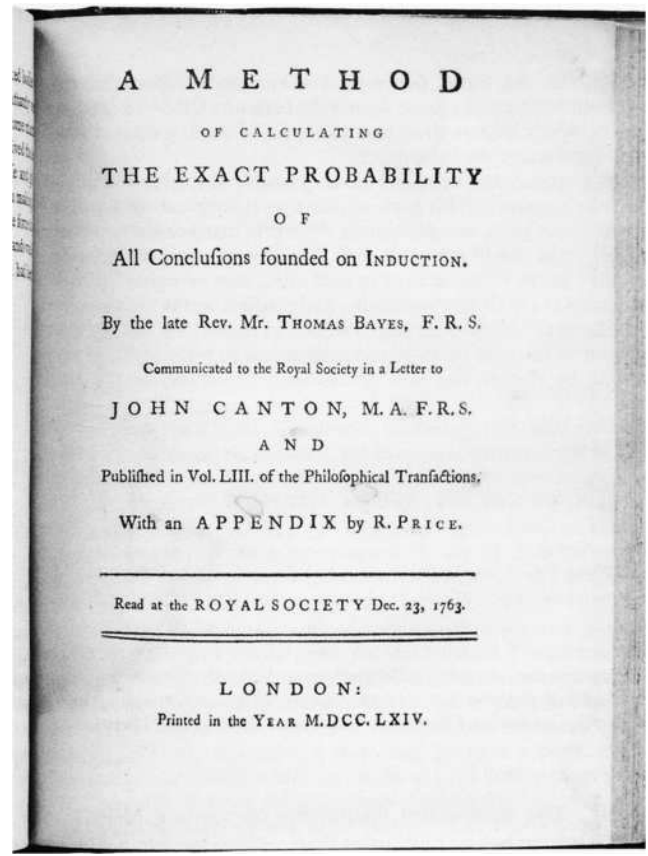}

\caption{The title page from the offprint of Bayes (\citeyear{Bay64}). Source: Watson (\citeyear{Wat}).}\label{fig3}
\end{figure}

\begin{figure}

\includegraphics{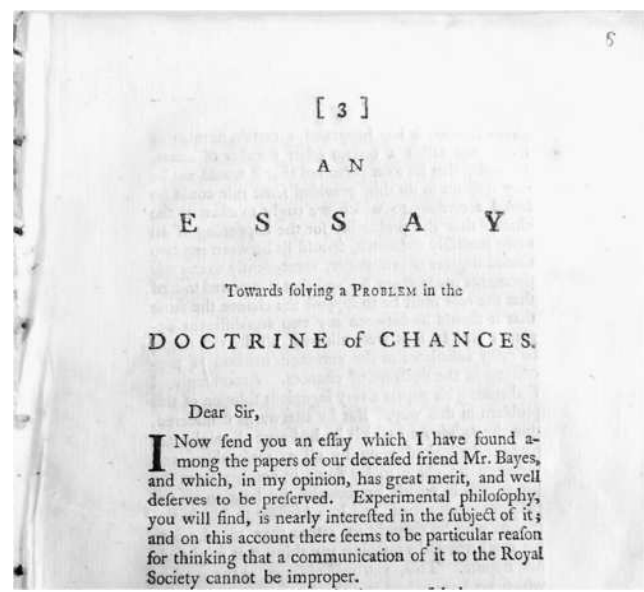}

\caption{Page 3 of the Bayes offprint, showing the journal title as a subtitle. Source: Beinecke
Rare Book and Manuscript Library, Yale University.}\label{fig4}
\end{figure}

\begin{figure}

\includegraphics{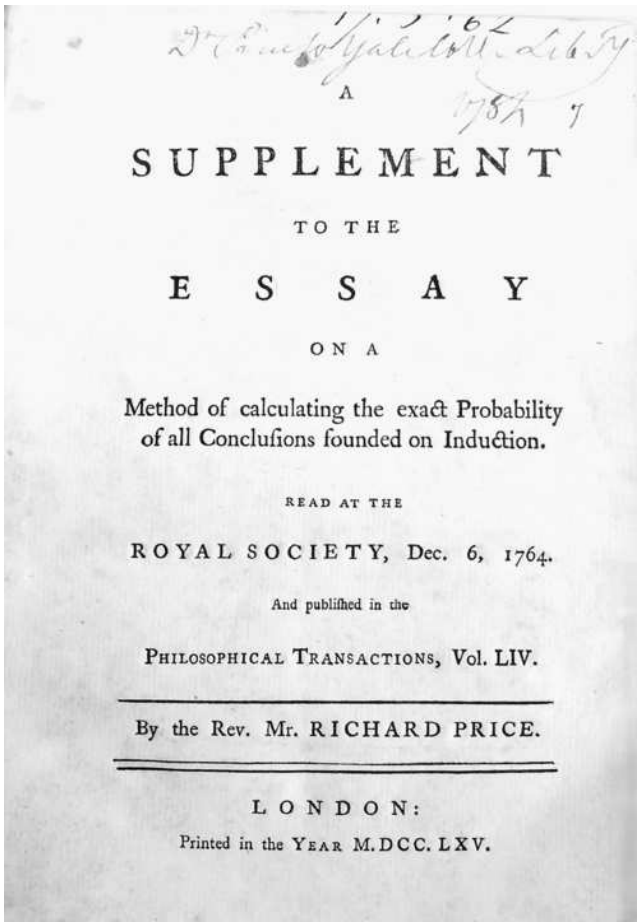}

\caption{The title page of the Price offprint. Source: Beinecke Rare Book and Manuscript Library,
Yale University.}\label{fig5}
\end{figure}

These are shown in Figures~\ref{fig3}--\ref{fig5}.

Where the commonly accepted title is almost completely uninformative, the offprint title is bold and
clear and promises even more than the paper delivers. This latter title surely originated from
Price, either as an afterthought or as a version omitted by the \textit{Transactions} editor as
too long or too bold. The offprint title clearly fixes the intention of the paper as addressing the
fundamental issue of induction, and it lends support to the following story of how it came to be
written and published.

\begin{longlist}[(1)]
\item[(1)] In 1748 David Hume published his famous essay ``Of Miracles'' (\cite{Hum}). The essay presented
his probabilistic argument for dismissing religious miracles, such as the story of Christ's
resurrection. Hume argued that the great improbability of the miracle (``a~violation of the laws of
nature'') overwhelmed the probability (far less than certainty) that the miracle was accurately
reported. Hume's essay caused quite a stir; it was widely read and much discussed and attacked.

\item[(2)] Thomas Bayes attempted to address Hume's argument, initiating a study of the application of
probability to induction in 1748 or 1749 with at least some of the calculations that were to appear
in the eventual paper. The earliest surviving notes of Bayes on probability contain these
calculations and have been dated to be prior to 31 December 1749 (Dale, \citeyear{Dal86}, \citeyear{Dal03}, page~429;
\cite*{Bel04}). In any event, Bayes put the work aside, possibly because he did not regard his
solution as mathematically fully satisfactory: his approximation to the incomplete beta function
that gave the posterior distribution was quite crude.

\item[(3)] At some point in 1749--1760 Bayes and Price met and discussed Hume's essay, with Price learning
of Bayes's mathematical work. We know they were close because Price was a beneficiary of a \pounds
100 bequest in Bayes's Will, drawn up in 1760. Both shared the same religion, the Dissenting Church;
both would have viewed Hume's essay as an assault on fundamental Church doctrine, as indeed Hume
intended it to be. Hume's essay would have had to come up in conversation. There is evidence
(discussed later) that Bayes told the philosopher David Hartley about his result in 1749; if Bayes
would share his work with Hartley, he would surely share it with a closer friend, Richard Price.

\item[(4)] Bayes died suddenly on 7 April 1761. Price, knowing of Bayes's work on this subject, sought out
his friend's manuscript and spent much of the next two years polishing and enlarging it for
publication. Price's goal in this---and this is the point that the offprint title introduces into
the theory---was from the beginning not simply to preserve a friend's work, but to wield it as a
formidable weapon against Hume's essay.

\item[(5)] The paper was read to the Royal Society on 23 December 1763 and published in 1764 both in the
journal and as an offprint. Price spent much of the next year trying to improve the accuracy of
Bayes's approximation to the incomplete beta integral that gives the posterior probability, reading
the result to the Royal Society on 6 December 1764, then publishing the work in 1765 both in the
\textit{Philosophical Transactions} and as an offprint.

\item[(6)] In 1767 the final event occurred, the deployment of this weapon. Price published the book
\textit{Four Dissertations}, explicitly taking on Hume on a number of fronts in four essays on
Providence. The fourth dissertation, ``On the Importance of Christianity, its Evidences, and the
Objections which have been made to it,'' was a direct probabilistic challenge to Hume's argument in
``Of miracles.'' The basic probabilistic point was that Hume underestimated the impact of there
being a number of independent witnesses to a miracle, and that Bayes's results showed how the
multiplication of even fallible evidence could overwhelm great improbability of an event and
establish it as fact (see \cite{Gil87}; \cite{Kru88}; \cite{DawGil89}; \cite{Ear02}; \cite{Zab05}). In his discussion Price referred to the paper using exactly the words in the offprint
title. Price included as a footnote, ``In an essay published in vol 53{rd} of the
\textit{Philosophical Transactions}, what is said here and in the last note, is proved by
mathematical demonstration, and a method shown of determining the exact probability of all
conclusions founded on induction,'' going on to present the results of the solution for a few cases
(\cite*{Pri67}, page~396). Price's footnote was quoted in full that same year in an anonymous review in
the \textit{Monthly Review, or, Literary Journal}, Vol. 36, for February 1767, page~90.
\end{longlist}

All of these facts individually, save the offprint title, have been well known for some time; the
offprint title permits assembling them in what seems to me a convincing narrative. The most
speculative step is the presumption that Bayes's own motivation was Hume's essay, since there is no
mention of this in the paper and most of Bayes's notes on this do not survive. But the dating fits,
and the discussion between Bayes and Price of this topic would have naturally occurred---Hume's
provocative essay was a major, widely discussed event in the philosophy of religion at exactly that
time. As \citet{Gil87} makes clear, the large number of responses to Hume signal that his essay had
``gone viral,'' to use a 21{st} century term. Hume visited Tunbridge Wells in 1756 while Bayes
was there, but it is not known if they met (\cite*{Dal03}, page~82).

As possible evidence that Bayes discussed his work with others, there is a passage in David
Hartley's \citeyear{Har} book, \textit{Observations on Man,} that sounds like Bayes's result and, indeed,
like the newly discovered offprint title. After mentioning De Moivre's direct result, what we now
call the weak law of large numbers for binomial trials, Hartley wrote,

\begin{quote}
``An ingenious Friend has communicated to me a Solution of the inverse Problem, in which he has
shewn what the Expectation is, when an Event has happened p times, and failed q times, that the
original Ratio of the Causes for the Happening or Failing of an Event should deviate in any given
degree from that of p to q. And it appears from this Solution, that where the Number of Trials is
very great, the Deviation must be inconsiderable: Which shews that we may hope to determine the
Propositions, and by degrees, the whole Nature, of unknown Causes, by a sufficient Observation of
their effects'' (\cite*{Har}, page~339).
\end{quote}

Bayes is one candidate for that ``ingenious Friend'' (Stigler, \citeyear{Sti83}, \citeyear{Sti99}, Chapter~15), and several
recent scholars have argued in his favor (\cite{Bel11}; \cite{Dal03}; \cite{Gil87}), in which case
the work was essentially complete within a year of Hume's publication on miracles.

This scenario would also provide an explanation for why Price seized the manuscript and dedicated so
much time to it. He was not Bayes's executor, and while one can imagine he might have been willing
to help publish a work after Bayes's death, his heroic efforts far exceeded the requirements unless
he had further, personal motivation. His adroit use of the essay in 1767 shows what that motivation
could have been. In 1815 Price's nephew, William Morgan, wrote that ``On the death of [Price's]
friend Mr. Bayes of Tunbridge Wells in the year 1761, [Price] was requested by the relatives of that
truly ingenious man, to examine the papers which he had written on different subjects, and which his
own modesty would never suffer him to make public.'' (Morgan, 1815, quoted in \cite*{Dal03}, page~259).
Morgan was writing at second hand more than a half-century after the event, and in any case the more
plausible scenario of Price initiating the contact and requesting permission to examine the papers
is consistent with what Morgan wrote.

Price must have circulated offprints to all those people he thought interested. Copies can be found
in a handful of research libraries today catalogued under the offprint title, including the
University of Edinburgh, the University of London and (missing the title page) Yale University. The
copy of the offprint offered in the catalogue \citet{Wat} apparently was sent by Price to his
friend Joseph Priestley, the discoverer of oxygen. (The asking price for a volume including both
offprints with others of less note was \pounds 45,000, and it sold promptly.) According to Thomas et
al. (\citeyear{ThoSteJon}, page~15), the printer's ledgers show that 50 copies of the Bayes offprint were produced
in June 1764.

Bayes's paper was almost universally ignored for more than a half-century following its publication,
and the uninformative nature of the title as first given must have played a role in this. Bayes was
mentioned briefly by Condorcet in 1781 (without giving a title) in the introduction to the volume of
the Paris academy with Laplace's second major memoir on inverse probability. Laplace himself briefly
mentioned Bayes, again without a title, in the historical section of his 1814 \textit{Essai
philosophique sur les probabilit\'{e}s}. Even in 1837, a~mention of the paper by S.-D. Poisson (\citeyear{Poi37}) was
very brief, giving no title and misspelling the author's name as
``Blayes.''\looseness=-1

In England the paper fared little better. It was ignored in 18{th} century encyclopedias, but
recognized reasonably well in a portion of Abraham Rees's \textit{Cyclopaedia} published in 1807.
There, an article on Chance gave what may be the first citation to the paper with the offprint
title, aside from that by Price himself and a mention in the list of Price's publications in
Priestley's \citeyear{Pri91} funeral discourse
(\cite{Pri91}).

\begin{quote}
``Among [the articles in the \textit{Philosophical Transactions}] which may be particularly
mentioned [is] an ``Essay on the Method of calculating the exact Probability\vadjust{\goodbreak} of all Conclusions
founded on Induction,'' and a ``Supplement'' to that essay:---the one preserved from the papers of
the late Rev. Mr. Bayes, and communicated, with an appendix, by Dr. Price, to the Royal Society in
the year 1762 [sic]; the other chiefly written by Dr. Price, and communicated in the following year.
These tracts contain the investigation of a problem, the converse of which had formerly exercised
the ingenuity of Mr. Bernoulli, De Moivre, and Simpson. Indeed both the problem and its converse may
justly be considered not only as the most difficult, but as the most important that can be proposed
on the subject; having (as Dr. Price well observes) ``no less an object in view than to shew what
reason we have for believing that there are in the constitution of things fixed laws, according to
which events happen; and that, therefore, the frame of the world must be the effect of the wisdom
and power of an intelligent cause; and thus to confirm the argument taken from final causes for the
existence of the Deity.'' (\cite*{Ree07}, page~3I:5--6.)
\end{quote}

This was likely written by William Morgan, who was credited with the article on Annuities in the
same \textit{Cyclopaedia} and who would have been expert in this area. He was also Richard
Price's nephew, and so this was not getting very far from the source.

Not all English sources were so appreciative. In 1809 an 18-volume abridgment of the
\textit{Philosophical Transactions} up to 1800 was published. In volume~12 Bayes's paper was
given a curt dismissal that entirely missed its originality:

\begin{quote}
LII. An Essay toward Solving a Problem in the Doctrine of Chances. By the late Rev. Mr. Bayes,
F.R.S. Communicated by Mr. Price. P. 370.

This problem is to this effect: ``Having given the number of times an unknown event has happened and
failed; to find the chance that the probability of its happening should lie somewhere between any
two named degrees of probability.'' In its extent and perfect mathematical solution, this problem is
much too long and intricate, to be at all materially and practically useful, and such as to
authorize the reprinting here;\vadjust{\goodbreak} especially as the solution of a kindred problem in Demoivre's
Doctrine of Chances, page~243, and the rules there given, may furnish a shorter way of solving the
problem. See also the demonstration of these rules at the end of Mr. Simpson's treatise on ``The
Nature and Laws of Chance'' (Hutton et al., \citeyear{HutShaPet09}, v.~12, page~41).
\end{quote}

However, the same Charles Hutton who helped compile these abridgments apparently took a different
view six years later. In 1815 he expanded the article on Chance in the first volume of the
2{nd} edition of his \textit{Philosophical and Mathematical Dictionary} (Bayes is not
mentioned in the first edition of 1795), by including the entire passage quoted above from Rees,
with citation to Rees (\cite*{Hut15}, v.~1, page~304).

By the twentieth century Price's preferred title had all but disappeared. Of the many historical
accounts that took Bayes seriously, including those cited in the references by \citet{Bel04},
Dale (\citeyear{Dal86}, \citeyear{Dal91}, \citeyear{Dal03}), \citet{Das88}, \citet{Edw92}, Hacking (\citeyear{Hac65}, \citeyear{Hac75}),
\citet{Hal98}, \citet{Pea78} and \citet{Sti86}, the closest to mentioning the title is Dale, who in an endnote mentions
that he learned from a University of Edinburgh librarian that the work catalogued there under the
offprint title was ``merely a reprint of the Essay'' (Dale, \citeyear{Dal91}, page~538). That the offprint title
was formally given as Price's preference seems to have escaped us all.

The published version of Bayes's Essay was prepared by Price as an edited and expanded version of
notes he had found in Bayes's papers, after Bayes died in April of 1761. Bayes's notes themselves
have been lost, save one small portion, and we do not know whether they even had a title, and if so,
what title. We also do not know whether Price's use of the uninformative title in the journal
publication was his choice or an editorial choice. We do know---as well as can be known---that
when it came time to construct an offprint title, the choice must have been his, for he would have
been paying the bill. Professor A. W. F. Edwards has pointed out to me that the offprint title is
the title given in the 10 volume set of Price's works published in 1815--1816.

The choice Price made for the offprint title does not directly come from the text of the printed
paper---similar words can be found there but none that make so all-encompassing a claim, a claim
that even a charitable reading of the paper would, strictly speaking, not support. But the new title
would better support the case against Hume, and Price may also have seen the need for a more
informative title, lest the work sink without a trace. Even with his bold choice, that is pretty
well what did happen, for it was only in the twentieth century that Bayes was, like the bones of an
ancient dinosaur, unearthed, dissected and put on prominent display for all to
admire.\looseness=-1

\section*{Acknowledgments}
I thank David Bellhouse, Andrew Dale, Lorraine Daston, Anthony Edwards, Dennis Lindley, Eugene
Seneta, Sandy Zabell and two referees for comments and additional references. It was Rick Watson's
scholarly catalogue that first brought the offprints to my attention; I have subsequently viewed
copies of both at Yale's Beinecke Library (where Price's Supplement is a copy presented to Yale
College in the 1780s by Richard Price himself, and Bayes's offprint now lacks the title page but may
have been received by the same route). I also inspected the copy of Price's offprint from Benjamin
Franklin's personal library, now held at the Library Company of Philadelphia as part of the
collection of the Historical Society of Pennsylvania. Franklin was a close friend of Price and he
may have had Bayes's offprint too at one point. A comparison of the two offprints with the journal
publications revealed no differences other than the resetting of part of the first journal page, the
addition of ``The End'' at the end, new pagination with the title page as number 1, different
printers' marks and, of course, the new title pages.



\end{document}